

\baselineskip=15pt plus 1pt minus 1pt
\tolerance=10000
\parskip=0pt
\parindent=15pt

\hoffset=.7truecm
\voffset=0.7truecm
\hsize=15truecm
\vsize=20truecm

\newdimen\myhsize   \myhsize=7.2truecm      
\newdimen\myvsize   \myvsize=40truecm      

\newdimen\pagewidth  \newdimen\pageheight
\pagewidth=\hsize  \pageheight=\vsize

\newinsert\margin
\dimen\margin=\maxdimen
\count\margin=0 \skip\margin=0pt

\def\myname{Yutaka Hosotani}
\def\mytitle{Anyons on a Torus}

\def\firstheadline{\hbox to \pagewidth{%
    {{\rm For the Proceedings of the 26th ICHEP} \hfil {\rm UMN-TH-1116/92}}%
       }}
\def\leftheadline{\hbox to \pagewidth{%
      {\it Page \folio\hss \myname\hss\qquad}%
      }}
\def\rightheadline{\hbox to \pagewidth{%
      {\it \qquad \hss\mytitle \hss Page \folio}%
       }}

\def\otherheadline{
   \ifodd\pageno \rightheadline \else \leftheadline \fi}


\def\onepageout#1{\shipout\vbox{
   \offinterlineskip
   \vbox to 1.truecm{    
      \ifnum\pageno=1 \firstheadline \else\otherheadline\fi \vfill}
   \vbox to \pageheight{
     \ifvoid\margin\else
       \rlap{\kern31pc\vbox to0pt{\kern4pt\box\margin \vss}}\fi
    #1
   \boxmaxdepth=\maxdepth} }
 \advancepageno}

\newbox\partialpage
\def\begindoublecolumns{\begingroup
  \output={\global\setbox\partialpage=\vbox{\unvbox255\bigskip}}\eject
  \output={\doublecolumnout} \hsize=\myhsize  \vsize=\myvsize }

\def\doublecolumnout{\splittopskip=\topskip \splitmaxdepth=\maxdepth
  \dimen1=\pageheight \advance\dimen1 by-\ht\partialpage
  \setbox0=\vsplit255 to\dimen1 \setbox2=\vsplit255 to\dimen1
  \onepageout\pagesofar \unvbox255 \penalty\outputpenalty}
\def\pagesofar{\unvbox\partialpage
  \wd0=\hsize \wd2=\hsize \hbox to\pagewidth{\box0\hfil\box2}}
\def\balancecolumns{\setbox0=\vbox{\unvbox255} \dimen1=\ht0
  \advance\dimen1 by\topskip \advance\dimen1 by-\baselineskip
  \divide\dimen1 by2 \splittopskip=\topskip
  {\vbadness=10000 \loop \global\setbox3=\copy0
    \global\setbox1=\vsplit3 to\dimen1
    \ifdim\ht3>\dimen1 \global\advance\dimen1 by1pt \repeat}
  \setbox0=\vbox to\dimen1{\unvbox1} \setbox2=\vbox to\dimen1{\unvbox3}
  \pagesofar}

\pageno=1

\font\bxBig=cmbx12  

\def\normal{\baselineskip=15pt plus 1pt minus 1pt}
\def\little{ \vskip 4pt}
\def\ep{\epsilon}
\def\eps{\varepsilon^{\mu\nu\rho}}
\def\d{\partial}
\def\la{\raise.16ex\hbox{$\langle$} }
\def\ra{\raise.16ex\hbox{$\rangle$} }
\def\go{\rightarrow}

\def\psibar{ \psi \kern-.65em\raise.6em\hbox{$-$} }
\def\Dbar{ D \kern-.8em\raise.65em\hbox{$-$} }

\def\N{ \kappa }

\def\x{ {\bf x} }
\def\y{ {\bf y} }

\def\r{ {\bf r} }

\def\jac{\vartheta_1}
\def\bjac{ {\overline{\vartheta}}_1}
\def\wbar{ w \kern-.7em\raise.4em\hbox{$-$} }

\baselineskip=7pt
{}~
\centerline{\bxBig ANYONS ON A TORUS}

\bigskip

\baselineskip=15pt

\centerline{\bf Yutaka Hosotani}
\centerline{\it School of Physics and Astronomy, University of Minnesota}
\centerline{\it Minneapolis, MN 55455, U.S.A.}
\centerline{and}
\centerline{\bf Choon-Lin Ho}
\centerline{\it Department of Physics, Tamkang University}
\centerline{\it Tamsui, Taiwan  25137, R.O.C.}

\bigskip
{\parindent=20pt
\centerline{Abstract}
\midinsert \narrower\narrower \noindent
We prove the equivalence between anyon quantum mechanics on a torus and
Chern-Simons gauge theory.  It is also shown that the Hamiltonian and total
momenta commute among themselves only in the physical Hilbert space.
\endinsert
}

\begindoublecolumns

\normal
\parindent=15pt

\centerline{\bf INTRODUCTION}
\little

A few years ago Einarsson$^1$ gave  braid group analysis of
 quantum mechanics of $q$ anyons
(with the statistics phase $\theta_s$) on a torus. He showed that Schr\"odinger
wavefunctions must have $M$-components, and that $q$, $M$, and $\theta_s$ must
satisfy
$$e^{2iq\theta_s} = 1 = e^{2iM\theta_s} ~. \eqno(1) $$
In particular, for $\theta_s=\pi/N$ ($N$: an integer), $q$ and $M$ must be
multiples of $N$.

On the other hand it is known$^2$ that anyon quantum mechanics on a plane is
equivalent to Chern-Simons gauge theory coupled to non-relativistic matter
fields.  Does the equivalence  remain valid on a torus?  If it does,  how
does the constraint (1) result from Chern-Simons gauge theory?

In this paper we show$^{3,4}$ that the equivalence is exact and everything
follows from Chern-Simons gauge theory.

\little
\centerline{\bf CHERN-SIMONS THEORY}
\little

The Lagrangian is given by
$$\eqalign{
{\cal L}&= {\N \over 4\pi} ~\eps a_\mu \d_\nu a_\rho  \cr
&+i \psi^\dagger  D_0 \psi
 - {1\over 2m} (D_k\psi)^\dagger (D_k\psi) ~~, \cr}  \eqno(2)  $$
where $D_0 = \d_0 +  i a_0$ and $D_k = \d_k - i a^k$.  The Chern-Simons
coefficient $\N$ is related to $\theta_s$ by $\theta_s=\pi/\N$.
To be definite, $\psi(x)$ is taken to be a fermion field.

On a torus (0$<$$x_j$$<$$L_j$, $j$=1,2) there are two non-integrable phases
of Wilson line integrals along non-contractible loops:
$$\exp \Big( i\int_{C_j} d\x \cdot {\bf a} \Big)
\go W_j=e^{i\theta_j} ~~. \eqno(3)   $$
They form a  conjugate pair$^{5}$:
$$[\theta_1, \theta_2] = {2\pi i\over \N} ~~. \eqno(4) $$
Fields are not single-valued in general:
$$\eqalign{
a_\mu[T_j x] &= a_\mu[x] + \d_\mu \beta_j(x) \cr
\psi[T_j x] &= e^{-i\beta_j(x)} \, \psi[x] \cr}   \eqno(5)  $$
where $T_1 \x= (x_1+L_1,x_2)$ etc.
Field operators must be smooth on the covering space, and therefore
$\psi[T_1T_2 x] = \psi[T_2 T_1x]$, from which the flux quantization condition
follows:
$$
\Phi = - \int d\x ~f_{12} =2\pi m
\hskip .2cm (m: ~{\rm integer}).    \eqno(6) $$

By solving Chern-Simons field equations $(\N/4\pi) \eps f_{\nu\rho}=j^\mu$,
$a_\mu(x)$ can be expressed in terms of $\theta_j(t)$ and $\psi(x)$.
The resulting Hamiltonian is
$$\eqalign{
&H = {1\over 2m}~\int d{\bf x} ~(D_k\psi)^\dagger (D_k\psi) ~~,\cr
&a^j(x) = {\theta_j(t)\over L_j} - {\Phi \over 2L_1L_2} \, \ep^{jk} x_k
  \, +  \cr
&\hskip .2cm  \int d\y \, \ep^{jk}\d^x_k G(\x-\y)
\Big( {2\pi\over \N} \psi^\dagger \psi(y) + {\Phi\over  L_1L_2} \Big)  \cr}
   \eqno(7) $$
where $G(\r)$ is the periodic Green's function on a torus satisfying
$\Delta G(\r) = \delta(\r) - (1/L_1L_2)$.
Furthermore one has to impose a constraint on physical states,
$$Q + {\N\over 2\pi} \Phi  \approx 0 ~~~
 \Big(  Q=\int d\x\, \psi^\dagger \psi \Big)   \eqno(8) $$
as (8),  despite being a part of the
original Chern-Simons field equations, does not follow from the Hamiltonian in
(7).

\little
\centerline{\bf VACUUM}
\little

The field theory defined by (7) and (8) with commutation relations for
$\theta_j$ and $\psi(x)$ is invariant under large gauge transformations:
$$\eqalign{
\theta_j~  &\go ~~ \theta_j + 2\pi n_j ~~, \cr
\psi(x) &\go e^{ 2\pi i  (n_1x_1/L_1 + n_2x_2/L_2) } ~\psi(x)  \cr}
  \eqno(9) $$
where ($n_1$,$n_2$)=(1,0) and (0,1).  The associated unitary operators are
given
by
$$\eqalign{
U_j =&\exp \bigg\{ i \ep^{jk}\, \N \, \theta_k   \cr
&\hskip 1.cm  - 2\pi i \int d\x \, {x_j\over L_j} \, \psi^\dagger \psi(x)
\bigg\}~. \cr}     \eqno(10) $$

$U_j$'s and $W_j$'s satisfy

$$\eqalign{
U_1\, U_2 \,&= \,U_2 \,U_1 ~ e^{-2\pi i \N} ~~,\cr
W_1 W_2 &= W_2 W_1 \, e^{-2\pi i/\N}~~.  \cr}   \eqno(11)  $$
Two gauge transformations $U_1$ and $U_2$ do not commute with each
other in general.$^{5,6}$   Consistent quantum theory is possible only if
the coeffiecient $\N$ is a rational number.$^{7}$   Two cases are important,
an integer $\N$  in the anyon superconductivity  and an inverse
integer $\N$ in the fractional quantum Hall effect.

Let us concentrate on the integer $\N$=$N$ case, in which $U_1$ and $U_2$
commute. As a consequence of (11) there are $N$ degenerate vacua.  Choosing
$U_j |0_a \ra = e^{i \alpha_j} |0_a \ra$, one finds
$$\eqalign{
W_1 |0_a\ra &=e^{-i\alpha_2/N} ~ |0_a \ra ~~,\cr
W_2 |0_a\ra &=e^{+i\alpha_1/N} |0_{a-1}\ra ~~.\cr}    \eqno(12)  $$

\little
\centerline{\bf WAVEFUNCTIONS}
\little
A $q$-particle Schr\"odinger wavefunction in quantum mechanics is a matrix
element of $q$ field operators $\psi(x)$ between the vacuum and
corresponding $q$-particle state $|\Psi_q \ra$.  One elaboration is necessary
on a torus.  It is given by
$$\eqalign{
\phi^{\rm f}_a&(t;\x_1, \cdots, \x_q)
= \la 0_a | \Omega \, \psi(1) \cdots \psi(q) | \Psi_q \ra ~~, \cr
\Omega &= \exp \bigg\{ - i \sum_{p=1}^q \Big( {x_1^p\over L_1} \theta_1
  + {x_2^p\over L_2} \theta_2 \Big) \bigg\} ~. \cr} \eqno(13) $$

A couple of things should be noted.
There are $N$ degenerate vacua so that the wavefunction must have
$N$-components: $\phi^{\rm f}_a$ ($a=1,\cdots,N$).  Secondly, the operator
$\Omega$ is necessary in the definition of $\phi^{\rm f}_a$ to make it
invariant under large gauge transformation (9).

Recalling $\theta_s$=$\pi/\N$,  we see that the constraint (1) is satisfied.
We have just shown that $M$=$N$, and (6) and (8) imply that $q$=$mN$.

\little
\centerline{\bf BRAID GROUP}
\little

The wavefunction $\phi^{\rm f}$ in (13) satisfies the braid group algebra
on a torus.   There are three sets of operations on a torus:  (a) $\sigma_j$:
the (counterclock-wise) interchange of the $j^{\rm th}$ and  ($j+1$)$^{\rm th}$
particles, (b)  $\tau_j$:  the loop transport of the $j^{\rm th}$ particle in
the $x_1$-direction, (c) $\rho_j$: the corresponding transport in the
$x_2$-direction.  These three, $\sigma_j$, $\tau_j$, and $\rho_j$ satisfy
the braid group algebra, from which Einarsson derived the aforementioned
constraint.$^1$

Action of these operators on $\phi^{\rm f}$ is simple:
$$\eqalign{
\sigma_j &: ~\x_j \leftrightarrow \x_{j+1}  \cr
\tau_j \,  &: ~\x_j \go T_1 \, \x_j \cr
\rho_j \,  &: ~\x_j \go T_2 \, \x_j  \cr}   \eqno(14)  $$
In particular, since $\psi$ is a fermion,
$$\sigma_j \, \phi^{\rm f} = -  \phi^{\rm f} ~~. \eqno(15)  $$
$\phi^{\rm f}$ is the wavefunction in the fermion representation.

As a consequence of (14), the braid group algebra is trivially satisfied.
However, $\phi^{\rm f}$ is tranformed quite nontrivially under the action
of $\tau_j$ and $\rho_j$:
$$\eqalign{
&( \tau_j \phi^{\rm f} )_a \cr
&={\rm exp} \bigg[ -i\beta_1(x^j)+ {i\pi\over N}\sum_p {x^p_2\over L_2}
+ {2\pi i a\over N} \bigg] \phi^{\rm f}_a \cr
&( \rho_j \phi^{\rm f} )_a \cr
&={\rm exp} \bigg[ -i\beta_2(x^j)-{i\pi\over N} \sum_p {x^p_1\over L_1}
   \bigg]   \phi^{\rm f}_{a-1}  \cr}
   \eqno(16) $$

Notice that $\phi^{\rm f}$ is a regular function of $\{ \x^p \}$, without
any singularity.  Under $\tau_j$ and $\rho_j$, $\phi^{\rm f}$ picks up
$\{ \x^p \}$ dependent phases, natural in gauge theory.   In Einarsson's
analysis it was implicitly assumed that phases must be constant, which demands
multi-valued wavefunctions.

\little
\centerline{\bf SINGULAR TRANSFORMATION}
\little

Einarsson's wavefunction, $\phi^{\rm E}$, is related to $\phi^{\rm f}$ by a
singular gauge transformation.$^{3,8,9}$
$$\eqalign{
&\phi_a^{\rm E} = \Omega_{\rm sing} \phi_a^{\rm f}   \cr
&\Omega_{\rm sing} = \prod_{j<k}
\bigg[ {\jac (w_{jk}) \over \bjac  (\wbar_{jk}) } \bigg]^{{1\over 2N}}
  \cdot e^{i\pi x_1^{jk}x_2^{jk} /L_1L_2}  \cr}
   \eqno(17)  $$
where $\jac(w)$ is Jacobi's theta function,
$x_1^{jk}=x_1^j -x_1^k$, $w=(x_1+ ix_2)/L_1$, etc..

It is straightforward to see that
$$\sigma_j \, \phi^{\rm E} = - e^{-i\pi/N} \, \phi^{\rm E} ~~.
  \eqno(18)  $$
The action of $\tau_j$ and $\rho_j$ is somewhat simplified.  The result
generalizes Einarsson's to arbitrary particle configurations.
All topological information is contained in $\Omega_{\rm sing}$.

\little
\centerline{\bf SCHR\"ODINGER EQUATION}
\little

The Schr\"odinger equation for $\phi^{\rm f}$
$$\eqalign{
i &{\d\over \d t}\, \phi_a^{\rm f} (t;\x_1,\cdots, x_q)
    = \hat H \, \phi_a^{\rm f} \cr
&=(q!)^{-1/2} \, \la 0_a | \Omega \,\psi(1) \cdots \psi(q) \, H | \Psi_q \ra
\cr}
   \eqno(19)  $$
is obtained by permuting $H$ (defined in(7)) to the left of $\Omega$ and
$\psi(j)$'s.  $\hat H$ is given by
$$\eqalign{
&\hat H = - {1\over 2m} \sum_{j}
    \big( \nabla^{(j)} - i {\bf A}^{(j)} \big)^2    \cr
&{\bf A}^{(j)k} = \ep^{kl} \, {2\pi\over N} \sum_{p\not= j}
 \Big( {x^j_l - x^p_l\over 2L_1L_2} \cr
&\hskip 2.5cm+  \nabla^{(j)}_l G(\x^j -\x^p) \Big)  . \cr}  \eqno(20)  $$

The equation for $\phi^{\rm E}$ is obtained by inserting (17) into (20).
The result is very simple:
$$i{\d\over \d t}\, \phi_a^{\rm E} = - {1\over 2m}
 \sum_j \big( \nabla^{(j)} \big)^2 ~ \phi^{\rm E}  ~~. \eqno(21) $$
It is a ``free'' equation.  The anyon interaction is hidden in the
boundary condition (18).

\little
\centerline{\bf TRANSLATION INVARIANCE}
\little

The total momentum operator in the second quantized theory is given by
$$P^k = -i \int d\x~ \psi^\dagger D_k \psi ~~. \eqno(22) $$
The corresponding operator in quantum mechanics is found to be
$$\hat P^k = -i \sum_j \nabla^{(j)}_k ~~, \eqno(23) $$
where $\hat P^k \phi_a^{\rm f} = (q!)^{-1/2} \la 0_a | ~\cdots~
 P^k |\Psi_q \ra$.

$P^k$ and $H$ in (7) form an algebra:
$$\eqalign{
[P^j, P^k] &= i\ep^{jk} {2\pi\over \N L_1L_2} Q\, \Big( Q+ {\N\over 2\pi}
\Phi \Big)  ~, \cr
[P^j, \,H\,] &= i\ep^{jk} {2\pi\over \N L_1L_2} J^k \Big( Q+ {\N\over 2\pi}
\Phi \Big)  ~, \cr} \eqno(24) $$
where $J^k$=$\int d\x\,j^k$ and =$P^k/m$ in the nonrelativistic theory.
They do not commute among themselves as operators, but do commute in the
physical Hilbert space defined by the constraint (8).

Hence the translation invariance is maintained in the Hilbert space.  Obviously
the corresponding operators in quantum mechanics, $\hat P^j$ and $\hat H$,
commute with each other.$^{10}$

Another important set of commutators are
$$\eqalign{
[W_j, P^k] &= \ep^{jk} {\pi\over \N L_1L_2} \{ \,Q\, , W_j \} ~, \cr
[W_j, \,H\,] &= \ep^{jk} {\pi\over \N L_1L_2} \{ J^k ,  W_j \} ~. \cr}
   \eqno(25)   $$
It turns out that the relations (24) and (25) remain valid even for
relativisitic theory with Dirac fields.  They are universal relations
in Chern-Simons theory.

\little
\centerline{\bf SUMMARY}
\little

Chern-Simons gauge theory was born  many years ago.  It is simple, beautiful,
and  rich.   It is important in  the fractional
quantum Hall effect and superconductivity.  It is the most powerful and
fruitful way of describing anyon physics, embodied with a unique algebraic
structure.  Much is hidden to be discovered in future.

\little
\centerline{\bf REFERENCES}
\little

\def\ap#1#2#3{{\it Ann.\ Phys.\ (N.Y.)} {\bf {#1}}, #3 (19{#2})}
\def\plB#1#2#3{{\it Phys.\ Lett.} {\bf {#1}B}, #3 (19{#2})}
\def\np#1#2#3{{\it Nucl.\ Phys.} {\bf B{#1}}, #3 (19{#2})}
\def\prl#1#2#3{{\it Phys.\ Rev.\ Lett.} {\bf #1}, #3 (19{#2})}
\def\prB#1#2#3{{\it Phys.\ Rev.} {\bf B{#1}}, #3 (19{#2})}
\def\prD#1#2#3{{\it Phys.\ Rev.} {\bf D{#1}}, #3 (19{#2})}

\parindent=18pt
\item{1.} T. Einarsson, \prl {64} {90} {1995}.

\item{2.}  R. Jackiw and S.Y. Pi, \prD {42} {90} {3500}.

\item{3.}  C.-L. Ho and Y. Hosotani,
UMN-TH-935/91, {\it Int.\ J.\ Mod.\ Phys.} {\bf A} (in press).

\item{4.}  C.-L. Ho and Y. Hosotani, TPI-MINN-92/41-T (Sep. 92).

\item{5.} Y. Hosotani, \prl {62} {89} {2785};  \prl {64} {90} {1691}.

\item{6.} X.G. Wen, \prB {40} {89} {7387}.

\item{7.}
K. Lee, Boston Univ. report,   BU/HEP-89-28;
A.P. Polychronakos, \ap{203}{90}{231};  \plB {241} {90} {37}.

\item{8.} S. Randjbar-Daemi, A. Salam, and J. Strathdee, \plB {240} {90} {121};
K. Lechner, Trieste SISSA preprint,  Thesis, Apr 1991.

\item{9.}
R. Iengo and K. Lechner, \np {346} {90} {551};  {\bf B364}, 551 (1991).

\item{10.}
R. Iengo, K. Lechner, and D. Li, \plB {269} {91} {109}.

\bye